# Experimental demonstration of a rapid sweep-tuned spectrum analyzer with temporal resolution based on a spin-torque nano-oscillator

A. Litvinenko[1,*], V. Iurchuk[1], P. Sethi[1], S. Louis[2], V. Tiberkevich[2], J. Li[2], A. Jenkins[3], R. Ferreira[3], B. Dieny[1], A. N. Slavin[2], U. Ebels[1]

[1]Univ. Grenoble Alpes, CEA, CNRS, Spintec, 38000 Grenoble, France

[2]Oakland University, 48309 Rochester MI, USA

[3]International Iberian Nanotechnology Laboratory (INL), 4715-330 Braga, Portugal

* corresponding author: artem.litvinenko@cea.fr

**It is demonstrated experimentally that a spin-torque nano-oscillator (STNO) rapidly sweep-tuned by a bias voltage can be used for *time-resolved* spectrum analysis of frequency-manipulated microwave signals with complicated multi-tone spectra. The critical reduction in the time of spectrum analysis comes from the naturally small time constants of a nano-sized STNO (1-100 ns). The demonstration is performed on a vortex-state STNO generating in a frequency range around 300 MHz, with frequency down-conversion and matched filtering used for signal-processing. It is shown that this STNO-based spectrum analyzer can perform analysis of multi-tone signals, and signals with rapidly changing frequency components with time resolution on a µs time scale and frequency resolution limited only by the "bandwidth" theorem. The proposed concept of rapid time-resolved spectrum analysis can be implemented with any type of micro and nano-scale frequency-swept oscillators having low time constants and high oscillation frequency.**

**Introduction**

In modern radar and communication technology it is important to be able to rapidly analyze the frequency composition of complicated external signals, and, in particular, to analyze frequency-agile signals, the frequency content of which is varying on the time scale comparable to the characteristic times of the signal propagation [1]. Some of the existing spectrum analyzers (SA) use a sweep-tuned-frequency approach to achieve a wide bandwidth of frequency analysis, high dynamic range, and low noise floor. They typically employ macroscopic voltage-controlled oscillators (VCO), such as CMOS and YIG-based oscillators that have sweep times down to 100 nanoseconds determined by the VCO time constants [2]. However, in certain radar and communication applications the spectrum analysis should be done even faster and, also, it is necessary to obtain the information about the temporal evolution of the spectra of frequency-agile signals with rapidly varying spectra.

Recent progress in nanotechnology have led to the development of *nano-sized oscillators*, called spin-torque nano-oscillators (STNO), whose time constants are naturally of the order of 1-100 nanoseconds [3]. The use of STNO in microwave spectrum analysis could thus bring at least a factor of 10 improvement to the time scale of spectrum analysis, and considerably impact radar and communication technologies.

The generation mechanism of STNOs is based on the effect of spin-transfer torque [4], that creates an effective negative damping in a DC current-driven magnetic tunnel junction structure, and that compensates the natural magnetic damping in the "free" magnetic layer of this structure [3, 4]. The frequencies of the periodic voltage signals generated by STNOs lie in the range from several hundred MHz to several tens of GHz [3-11]. Most importantly, the STNO frequency strongly depends on the oscillation amplitude, and, therefore, on the amplitude of the bias direct current driving the STNO. This non-isochronous property of STNOs creates the possibility to modulate the generation frequency of an STNO by modulating the magnitude of the driving bias direct current (or voltage) [12-15]. Theory [5-6] predicted that the maximum speed, at which the STNO frequency can be modulated by changing the current, is given by the relaxation rate $\Gamma_p = \pi f_p$. This is the rate with which fluctuations of the amplitude of an STNO-generated signal relax to a quasi-stationary limit cycle. Experiments have shown that characteristic frequencies $f_p$ of this relaxation process depend on the STNO configuration and range from $f_p$=1-10 MHz for vortex-state STNOs [16] to $f_p$=100-400 MHz for uniform magnetized STNOs [15, 17-19]. However, it was also predicted [20] and experimentally [21, 22] demonstrated that using field modulation it is possible to achieve modulation frequencies on the order of GHz. Hence when using a "sawtooth" modulation signal (either current or field with frequency $f_{sw}$) one can expect to sweep the STNO frequency without significant distortions within sweep times $T = 1/f_{sw}$ of at least one microsecond for vortex based STNOs but that could potentially be as fast as 1-10 ns for other STNO configurations.

In a recent theoretical paper [23] it was proposed to use such a sweep-tuned STNO as a central element of a novel ultra-fast spectrum analyzer where signal mixing and correlation processing using low-pass and matched filters were involved.

Here we demonstrate a proof-of-principle experimental realization of this method of fast spectrum analysis [23] for the case of vortex-based magnetic tunnel junction STNOs generating stable sinusoidal signals in the 280-320 MHz frequency range and characterized by an amplitude relaxation frequency of $f_p$= 2-3 MHz. The largest frequency sweeping rates $\rho_{sw}$ in our experiments were exceeding $\rho_{sw} = \Delta F_{sw}/T = 30$ MHz/µs for a frequency span of $\Delta F_{sw}$ = 20 MHz, allowing us to achieve spectrum analysis of complex multi-tone and frequency manipulated signals with temporal resolution on the µs time scale.

## Results

The scheme of the experimental STNO-based device used to perform spectrum analysis in our experiments is shown in Fig. 1. It consists of two blocks: the block (a) containing an STNO and used for generation of a sweep-tuned reference signal $V_{ref}$, and the block (b) used for signal processing and eventual spectrum analysis of an external signal $V_{in}$.

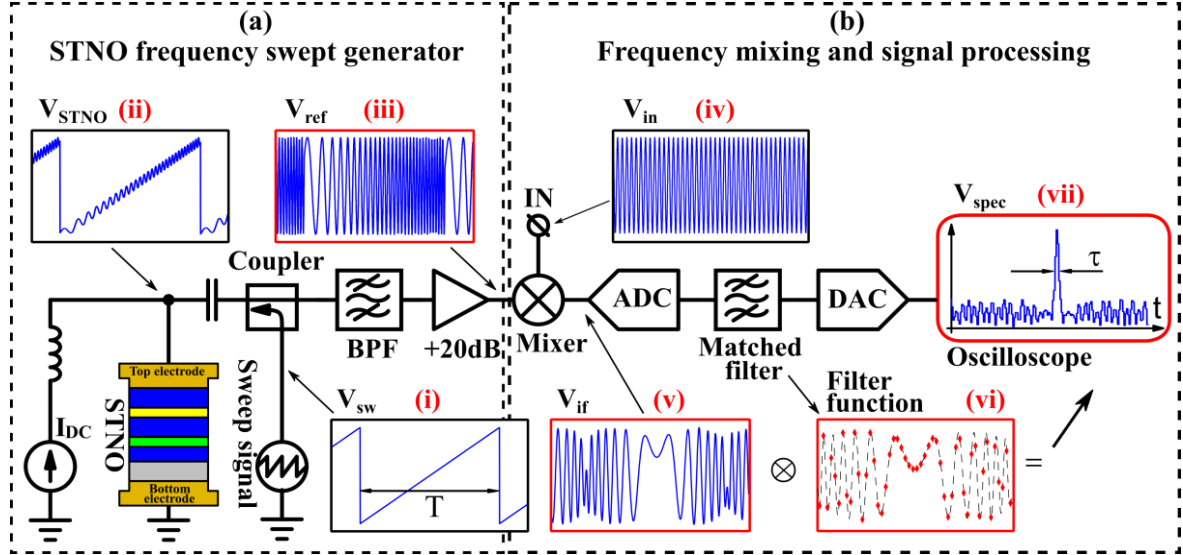

**Figure 1** (two column figure): Schematic of the STNO-based spectrum analyzer consisting of two blocks: block (a) generating a sweep-tuned reference signal $V_{ref}$, and block (b) performing signal processing resulting in the spectral analysis of the incoming signal $V_{in}$. The insets show the voltage signals vs. time at different points of the spectrum analyzer circuit: (i) the sweep signal $V_{sw}$, (ii) the output of the STNO $V_{STNO}$, (iii) the reference signal $V_{ref}$ produced by the sweep-tuned STNO after filtering and amplification, (iv) the external signal $V_{in}$, (v) the mixer output signal $V_{if}$, (vi) the discretized basis filter function of the matched filter signal, (vii) the output signal $V_{spec}$ resulting in a peak of duration $\tau$ and containing the information about the spectrum of the external signal $V_{in}$.

The sweep-tuned reference signal $V_{ref}$ is produced as follows. First, a constant DC current $I_{DC}$ is applied to the STNO to generate a sinusoidal auto-oscillation signal of a constant carrier frequency $f_0$. Then, an additional voltage $V_{sw}$ having a "saw-tooth" shape in time and a frequency $f_{sw}=1/T$ is applied to the STNO via a coupler. The amplitude of $V_{sw}$ is adjusted in such a way, that the frequency of the auto-oscillation signal generated by the STNO sweeps between two values $f_1$ and $f_2$, around $f_0$ with a span $\Delta F_{sw}=f_2 - f_1$. After passing through a band-pass filter and an amplifier, the reference signal, $V_{\text{ref}}$, has a frequency chirp with the instantaneous frequency $f_{\text{ref}}(t)$ that changes in time.

For the analysis of an external signal $V_{\text{in}}$ of an unknown frequency $f_{\text{in}}$, the reference signal $V_{\text{ref}}$, along with the external signal $V_{\text{in}}$, are supplied to the mixer situated in the block (b). For signal mixing we used a commercially available mixer (AD831) having an output cutoff frequency of $F_F =200$ MHz, and, therefore performing the function of a low-pass filter with a bandwidth of $\Delta F_F =200$ MHz. We note, that in our experiments the sweeping frequency interval $\Delta F_{sw}= 20$ MHz was much smaller than the mixer bandwidth of $\Delta F_F$, so that the frequency resolution of the spectrum analysis was determined only by the sweeping frequency $f_{\text{sw}}$.

The output signal of the mixer $V_{\text{if}}$ has the intermediate frequency $f_{\text{if}}(t)$ equal to the difference of frequencies of the mixed signals $f_{\text{if}}(t) = f_{\text{ref}}(t)- f_{\text{in}}$. This instantaneous frequency goes to zero at a certain time $t_0$, when $f_{\text{ref}}(t_0) = f_{\text{in}}$. Note, that the component of the mixed signal having the frequency $f_{\text{ref}}(t) + f_{\text{in}}$ was removed due to the $F_F$ output cutoff frequency of the mixer.

The output signal $V_{\text{if}}$ of the mixer is then digitized using an 8-bit AD9280 ADC, and passed through a matched filter, implemented with a Field Programmable Gate Array (FPGA), to compress this signal into a narrow peak of duration $\tau$ in $V_{\text{spec}}$. The matched filter output is then converted back into the analog domain using an 8-bit resolution AD9708 DAC, and is visualized on a single shot oscilloscope.

The temporal position of the maximum of the peak in $V_{\text{spec}}$ determines the frequency of the input signal $f_{\text{in}}$, while the temporal width of this peak $\tau$ characterizes the resolution bandwidth (RBW), or accuracy of the frequency analysis:

$$\Delta F_{RBW} = \rho\tau = \Delta F_{sw}\frac{\tau}{T}. \qquad (1)$$

We used in our signal processing protocol a *matched filter* instead of a conventional band-pass filter with envelope detection in order to achieve a resultant peak having a single maximum and a minimum possible duration $\tau$, which is independent of the phase difference between the input signal $V_{in}$ and the STNO-generated reference signal $V_{ref}$ (see [23] for details). Details on the matched filter configuration and implementation can be found in the Method and Supplementary Materials sections.

To demonstrate experimentally the STNO-based spectrum analysis we have chosen a relatively low frequency (~ 300 MHz) vortex-state STNO as STNOs of this type have rather large output power of order of µW and typical linewidth of a few hundred kHz [16, 24, 25]. The results are presented for two different devices, although many other devices demonstrated very similar results (see "Methods" section for details on the devices and their fabrication).

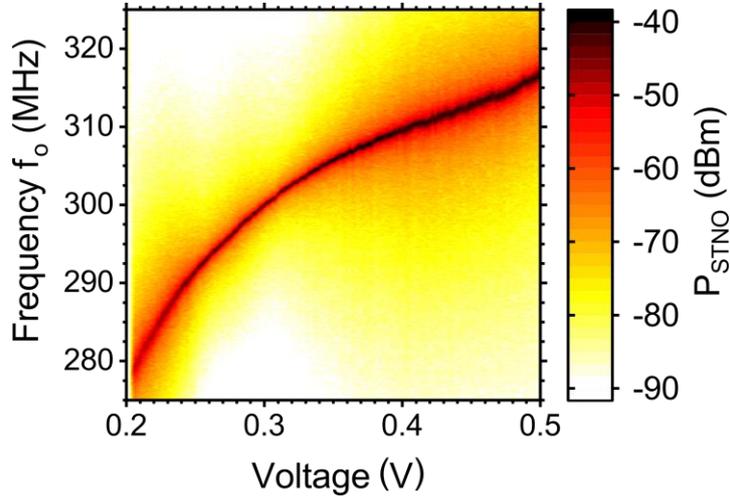

Figure 2 (one column figure): Nonlinear frequency-voltage characteristics $f_o(V)$ of the vortex-state STNO used in our experiments, performed under an out-of-plane bias magnetic field of $H_\perp$=3 kOe.

Fig. 2 shows a free-running frequency-voltage characteristic $f_o(V)$ of a representative vortex-state STNO device obtained under the action of an applied out-of-plane bias magnetic field $H_\perp$=3 kOe. It is clear from Fig. 2 that when the DC voltage applied to the STNO is varied between 0.23 V and 0.37 V, the frequency generated by the STNO is a *non-linear* (convex) function of voltage, and varies between $f_1$=287 MHz and $f_2$=307 MHz ($\Delta F_{sw}$ =20 MHz) with a typical linewidth (full width at half maximum (FWHM)) of its power spectrum equal to $2\Delta F = 0.25$ MHz.

While this nonlinearity of the frequency variation with voltage does not create serious difficulties in the determination of the frequency of *monochromatic* external signals lying in the center of the analyzed frequency range, it substantially deteriorates the frequency analysis of signals with complicated multi-tone spectra distributed over the whole analyzed frequency range. This difficulty can be eliminated by using a nonlinear (concave) profile of a "quasi-sawtooth" voltage applied to the STNO, which makes it possible to obtain a purely linear time dependence of the chirped frequency in the signal $V_{ref}$ generated by the sweep-tuned STNO (see Supplementary Materials for details).

This method of compensation of the nonlinearity of the STNO frequency-voltage characteristic $f_o(V)$ was used for the analysis of signals having rapidly changing frequency components (see Fig. 5c,d), while for the analysis in Figs. 3, 4, 5 a,b a linear "saw-tooth" voltage $V_{sw}$ was used.

First, we present the results of the analysis of a monochromatic sinusoidal signal $V_{in}$ to demonstrate that (i) using a vortex STNO-based spectrum analyzer, one can achieve frequency sweep rates of the order of MHz/μs and sweeping frequencies of the order of MHz and (ii) the resolution in external frequency determination is close to the theoretical limit for fast spectrum analysis. Then, we demonstrate that the STNO-based spectrum analyzer is capable of performing the analysis of signals having a complex frequency spectrum, such as two-tone signals, and signals with time-varying frequency components.

The experimental results are shown in Fig. 3 demonstrating the analysis of a monochromatic (single-tone) sinusoidal signal ($f_{in}$ = 300 MHz, $P_{in}$ = 1.0 mW) supplied to the spectrum analyzer from a commercial signal generator. The three panels in Fig. 3 show : (top) the sweeping "sawtooth" voltage $V_{sw}$ having the frequency $f_{sw} = 1/T = 0.4$ MHz; (middle) the chirped-frequency signal $V_{if}$ obtained after the mixer; (bottom) the output signal from the matched filter $V_{spec}$ measured using a standard oscilloscope.

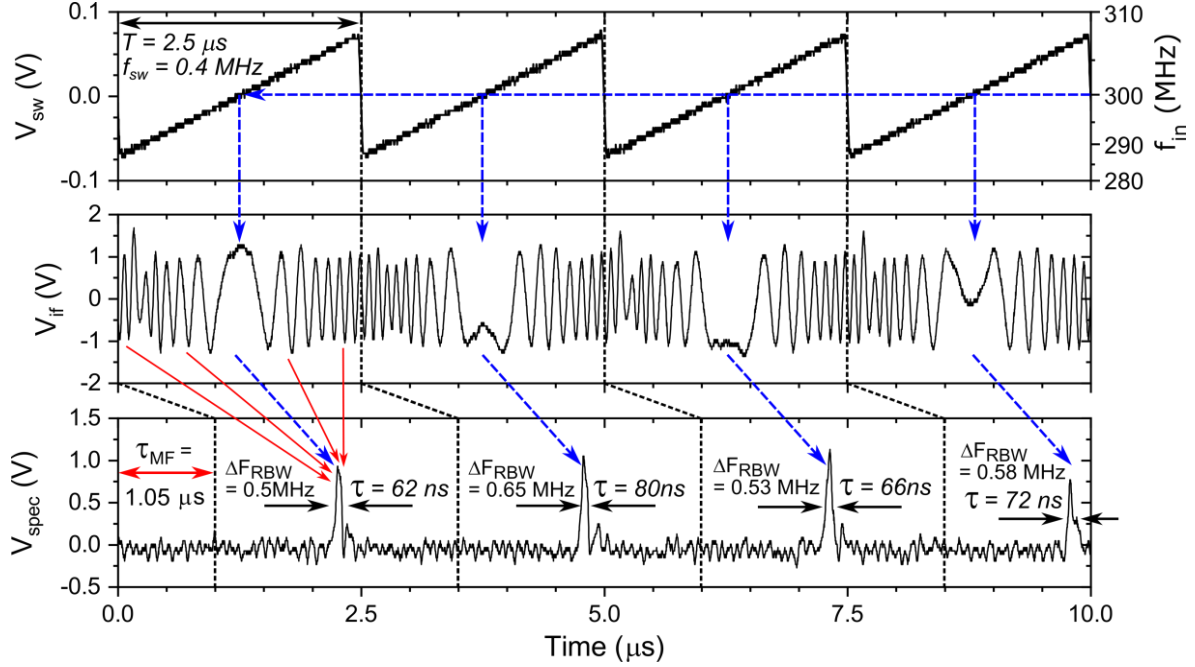

Figure 3 (two column figure): Spectral analysis of a single-tone external sinusoidal signal ($f_{in}$ = 300 MHz, $P_{in}$ = 1.0 mW): sweeping voltage signal $V_{sw}$ (top panel), $V_{if}$ (middle panel) and $V_{spec}$ (bottom panel) for $f_{sw}$ = 0.4 MHz taken at different points of the spectrum analyzer setup (see insets (i), (v) and (vii) of Fig. 1). The experimental frequency resolution $\Delta F_{RBW}$ was calculated using (1) from the experimentally measured peak duration $\tau$ (see bottom panel).

At the time $t_0 \approx 1.25$ µs (corresponding to $V_{sw} = 0$) the STNO frequency $f_{ref}(t_0)$ is the same as the frequency of the external signal $f_{in}$ = 300 MHz, resulting in a frequency difference $f_{if}(t_0) = f_{ref}(t_0) - f_{in} = 0$. It is this moment which has to be identified and measured. The chirped signal $V_{if}$ after passing through a matched filter is squeezed in time, and forms in the output signal $V_{spec}$ a pronounced peak of the duration $\tau$ at $t = t_0 + \tau_{MF}$, where the $\tau_{MF}$ is the phase delay in the matched filter and is always constant for any input chirped signal. Taking into account the phase delay in the matched filter one can thus correctly identify the frequency of the input signal (see bottom panel and the corresponding frequency scale to the right in the top panel).

The accuracy of the frequency determination using the STNO-based spectrum analyzer, characterized by the frequency resolution bandwidth $\Delta F_{RBW}$, can be evaluated from the experimentally measured magnitude of the peak duration $\tau$ using Eq. (1) (see Fig. 4).

In Fig. 4(a) we present the results of the spectrum analysis ($V_{spec}(t)$) of a monochromatic external signal for three different sweeping frequencies $f_{sw}$= 0.1, 0.8, and 1.5 MHz, while in Fig. 4(b) we compare the experimentally obtained frequency resolution bandwidth $\Delta F_{RBW}$ with a theoretical limit obtained from the "bandwidth" theorem $\Delta F_{RBW} = f_{sw} = 1/T$.

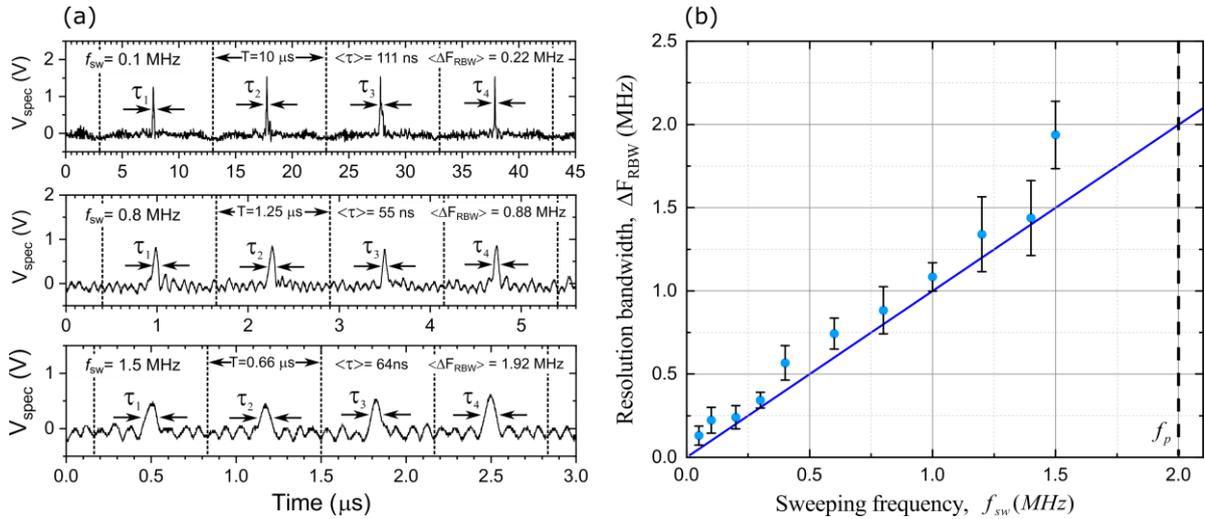

Figure 4 (two column figure): (a) Results of the spectrum analysis ($V_{spec}(t)$) of a monochromatic external signal for three different sweeping frequencies $f_{sw}$ =0.1 , 0.8, and 1.5 MHz; (b) Resolution bandwidth (RBW) of the spectrum analysis $\Delta F_{RBW}$ as a function of the sweeping frequency $f_{sw}$. The values of corresponding to the light blue dots presented in Fig. 4 (b) were calculated from the measured values of the resulting peak duration $\tau$ using Eq.(1). Error analysis was done on four random consequent measurements of the peak duration $\tau$. The straight dark blue line in Fig. 4(b) was calculated from the "bandwidth" theorem: $\Delta F_{RBW} = f_{sw} = 1/T$. The dashed vertical line shows the value of the experimental vortex STNO relaxation frequency $f_p$ for amplitude fluctuations limiting the upper modulation frequency of the STNO-generated signal (see [5, 15, 16] for details).

It is clear, that for the smaller sweeping frequencies the frequency resolution bandwidth $\Delta F_{RBW}$ of the spectrum analysis will be limited from below by the STNO generation linewidth $2\Delta F = 0.25$ MHz [26]. However, as it can be seen from Fig. 4, for sufficiently large sweeping frequencies, $f_{sw} > 0.1$ MHz ($\rho_{sw} > 2$ MHz /µs), the experimentally obtained frequency resolution bandwidth $\Delta F_{RBW}$ is very close to the theoretical limit defined by the "bandwidth" theorem, which means that a large generation linewidth of an STNO and its relatively high phase noise have practically no influence on the resolution bandwidth $\Delta F_{RBW}$ in the proposed method of rapid spectrum analysis.

The results presented so far demonstrated the basic properties of the STNO-based spectrum analyzers when working with simple monochromatic external signals. Below, we demonstrate that STNO-based spectrum analyzers can work successfully with rather complex external signals having many different frequency components that can vary in time.

The first example is given in Fig. 5a where we demonstrate spectrum analysis of an external signal that is a superposition of two single-tone continuous signals supplied from two commercial signal generators with frequencies $f_{in1} = 300$ MHz, and $f_{in2} = 306$ MHz. The top panel in Fig. 5a shows the linear voltage sweep $V_{sw}$ at $f_{sw} = 0.3$ MHz, while the vertical dashed lines show the positions where the sweeping STNO frequency equals the frequency components contained in the external signal. The bottom panel of Fig. 5a shows the resultant voltage $V_{spec}$ at the output of the matched filter with two distinct peaks, which occur at times corresponding to the frequency components $f_{in1}$ and $f_{in2}$ in the input signal. Note, that the time shift between the linear sweep voltage $V_{sw}$ and the resultant voltage $V_{spec}$ at the output of the matched filter (see Fig. 3) caused by the time delay $\tau_{MF}$ in the matched filter was not shown in Fig. 5(a,b,c).

Fig. 5(b) shows the result of the frequency analysis of an external signal whose frequency is hopping between the values of $f_{in1} = 301$ MHz and $f_{in2} = 299$ MHz. This experiment demonstrates that the STNO-based spectrum analyzer can easily detect the changes of external frequency in time if these changes are happening on the time scale that is larger than the period $T$ of the STNO frequency sweep. This property will be very important to efficiently analyze signals of modern radar using frequency hopping protocols.

The ultimate demonstration of the potential of the STNO-based sweep-tuned spectrum analyzer is given in Fig. 5(c), (d) where the analysis of complex external signals whose frequency is continuously or discontinuously changing in time is illustrated. If such changes are happening on the time scale $T_n = nT$ that is $n$ times larger, than the STNO sweeping period $T$, the STNO-based spectrum analyzer successfully tracks the temporal evolution of the frequency of the external signal. In particular, in Fig. 5c the frequency of the external signal varies in time between $f_{min}$ = 295 MHz and $f_{max}$ =305 MHz in a sinusoidal fashion with the period $T_{10} = 10T$ ($f_{sw}$= 1/T =0.5 MHz).

In such a case to recover the true temporal frequency variation in the external signal we performed the compensation of nonlinearity of the STNO characteristics $f_o(V)$ Fig. 2 (see Supplementary Materials for details), and used a non-linear (concave) sweeping voltage signal $V_{sw}$ (see the top panel in Fig. 5c). The peaks in the signal $V_{spec}$ having different temporal positions corresponding to different instantaneous frequencies of the external signal measured during consequent sweeping periods are shown in the middle panel of Fig. 5c, while the resultant temporal evolution of the frequency of the external signal $f_{in}(t)$ is shown in the bottom panel of Fig. 5c. In this panel the frequency axis is simply the corresponding frequency scale of a single sweeping period of the middle frame of the same figure.

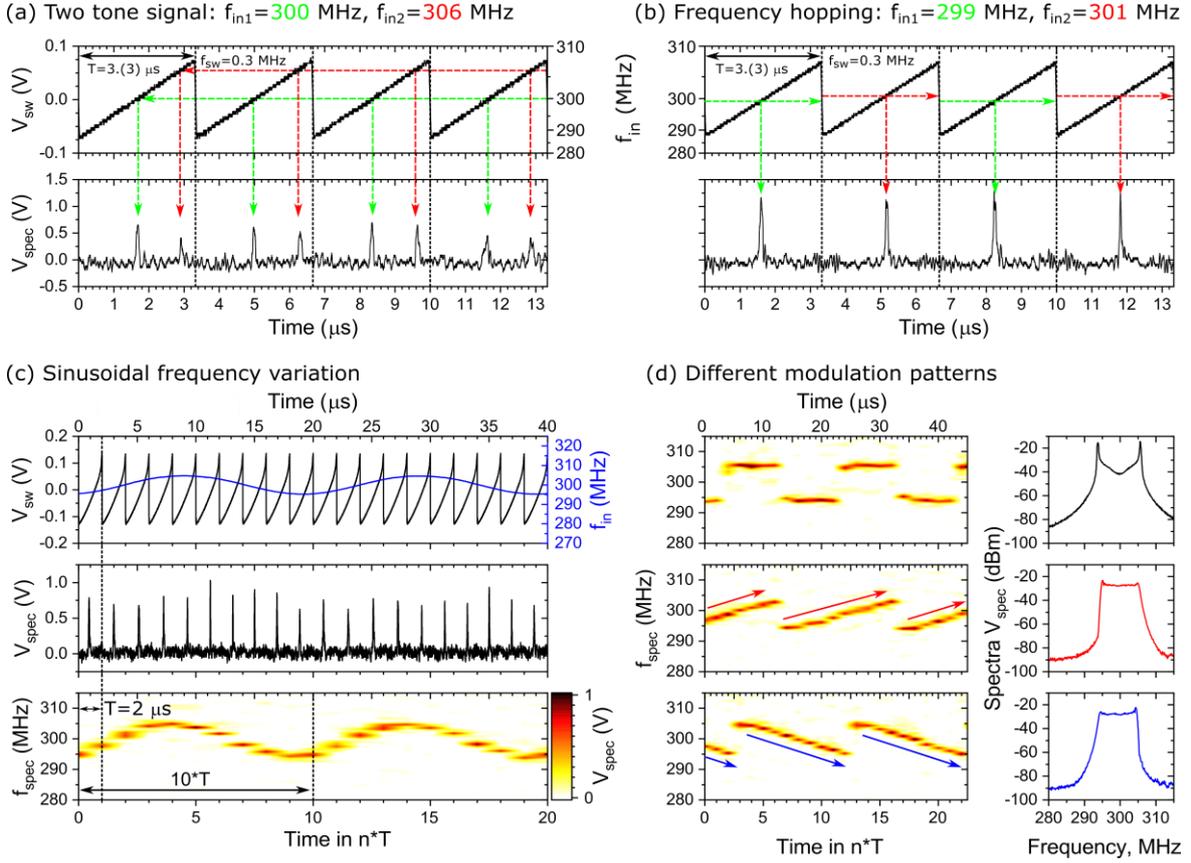

Figure 5 (two column figure): Spectrum analysis of complex external input signals $V_{in}$: (a) Two-tone signal containing frequency components $f_{in1}$=300 MHz and $f_{in2}$=306 MHz; (b) Signal with a frequency that is hopping between two close values of $f_{in1}$=301 MHz and $f_{in2}$=299 MHz ; (c) Signal with a frequency that is varying continuously in a sinusoidal fashion with the period $T_{10}$=10 $T$, where $T$ is the sweeping period. Top panel shows the nonlinear sweeping voltage $V_{sw}$, middle panel shows the resultant peaks $V_{spec}$ with temporal positions varying within each sweeping period, while the bottom panel shows the experimentally obtained variation of the external signal frequency; (d) Signals with various modulation patterns with the period $T_{10}$=10$T$=20 $\mu s$ and arbitrary initial phases of modulation. Top panel shows the signal with the shift-keyed frequency between 295 and 305 MHz. Middle and bottom panels show the signals whose frequency is varying discontinuously in a "saw-tooth" fashion with increasing and decreasing frequency, respectively. The frames at the right show the results of a conventional spectrum analysis for the same signals.

Finally in the three panels of Fig. 5d we show a similar result of the analysis of input signals where the frequency experiences discontinuous shift keying (top panel) and saw-tooth-like temporal variations with either increasing (middle panel) or decreasing (bottom panel) frequency. It is clear, that in a sweep-tuned spectrum analysis based on an STNO it is possible to reveal and analyze any fast-changing modulation patterns, while the power spectra obtained with a conventional frequency-swept spectrum analyzer (shown in panels to the right in Fig. 5d) for saw-tooth signals with increasing and decreasing frequencies are very similar, and demonstrate only the same interval of frequency variation in both analyzed signals but do not provide any time resolution for the changing frequency of the incoming signal.

**Conclusion**

In this work, a novel approach to time-resolved high-speed spectrum analysis was experimentally demonstrated. The key element used in this method is a nano-sized active STNO device that plays the role of a sweep-tuned oscillator. The nano-scale size brings the advantage of short intrinsic time constants that, together with the excellent tunability of STNOs, allows to rapidly tune the STNO frequency in response to a fast-changing control signal.

The repetition of the fast tuning of the STNO through the studied frequency range brings the capability of temporal resolution in the spectral analysis of the frequency-agile signals.

The proof-of-principle experimental demonstration for this STNO-based spectrum analysis was performed for a vortex-state STNO characterized by generation frequencies around 300 MHz with a maximum modulation or sweep frequency of $f_{sw}^{max}$=1.5 MHz, and the shortest time interval of spectrum analysis $T_{min} = 0.67$ µs.

Besides single tone (or monochromatic signals), we also demonstrated the possibility of time-resolved spectrum analysis of complex dynamically modulated signals (see Fig. 5), that is not possible to achieve using conventional spectrum analyzers. Furthermore, with the matched filter technique for signal processing the experimentally determined frequency resolution bandwidth $\Delta F_{RBW}$ (see Fig. 4 b) follows the "bandwidth" theorem. An important result here is that at high sweeping frequencies (and high sweep rates) the frequency resolution bandwidth $\Delta F_{RBW}$ becomes independent of the STNO phase noise and linewidth, and is given only by the sweeping frequency $f_{sw}$ and the experimental setup (filters).

The proof-of-concept demonstration for this novel approach to rapid and time-resolved spectrum analysis was done here for a vortex STNO demonstrating relatively low generation frequencies, and, consequently, low modulation rates. Our theoretical estimations [5, 23] show, and existing experimental results confirm, that STNOs having a uniformly magnetized "free" layer can generate frequencies $f_o$ up to several tens of GHz [3, 7-11] and can have modulation frequencies reaching several hundred MHz [15, 19] and up to GHz [20-22]. Thus, the STNO-based ultra-fast time-resolved spectrum analysis with high frequency resolution is a solvable engineering problem. We firmly believe that the ultra-fast STNO-based spectrum analyzers with nanosecond time resolution will become in the near future practical and highly competitive microwave signal processing devices.

**Methods**

**MTJ devices**

The STNOs used for the experiments are magnetic tunnel junction nanopillar devices with a vortex-state configuration of the free layer. The devices, realized at the International Iberian Nanotechnology Laboratory (INL), have the following composition (thicknesses in nm): substrate/ 6 IrMn/ 2.6 CoFe30/ 0.85 Ru/ 1.8 CoFe40B20/ MgO/ 2.0 CoFe40B20/ 0.2 Ta/ 7 NiFe/ 10 Ta. The materials were sputter deposited using a Singulus Rotaris machine and were nanofabricated using ion beam and chemical etching techniques. Best signal generation properties in terms of phase noise and frequency tuning were obtained for nanopillar diameters of 300-400nm. The corresponding frequencies lie in the range of 270-330 MHz when applying an out-of-plane field $H_\perp$ at a small in-plane tilt angle of 1-5°. For the results presented in the manuscript we used a permanent magnet at variable distance to produce an out-of-plane field of $H_\perp \approx 3$kOe. Results are presented for devices that had a diameter of 370 nm, a parallel resistance of $R_{STO} \approx 40\Omega$, a TMR≈150% and an RA≈3 $\Omega$ µm².

**Experimental setup**

**Free running STNO**: The characterization of the free running signal was done with the setup of Fig. 1, where $V_{sw}$ was set to zero and the STNO signal $V_{STNO}$ after an amplification of 20dB was analyzed using a spectrum analyzer or a single shot oscilloscope. From the spectrum analyzer measurements we obtain the frequency $f_o$, full width half maximum linewidth and output power of the STNO signal after subtraction of the measurement gain. Representative results are shown in the Supplementary Material Fig. S1. From the time traces registered on the oscilloscope we extract the instantaneous frequency and phase of the STNO using the method of the Hilbert transform [16, 17, 18]. This method was used to determine the amplitude relaxation frequency [16, 18], the instantaneous STNO frequency when applying the sweep voltage $V_{sw}$ and to verify the compensation of the frequency – voltage nonlinearity (Fig. S2 in Supplementary Material section II).

**SA**: The spectrum analyzer setup described in Fig. 1 differs slightly from the one proposed in Ref. 23 where the mixing of the reference signal $f_{ref}(t)$ and the input signal $f_{in}$ was proposed to occur within the STNO, without an external mixer. Since the frequency range of the mixed signal $V_{if}$ is close to the frequency range of $V_{sw}$ and therefore the ranges may overlap, it is not always possible to well separate these signals using filters. Therefore, $V_{sw}$ was removed from $V_{STNO}$ before mixing using a band pass filter with a range of 250-350MHz and the mixing was achieved using an external mixer.

**Matched Filter**

The matched filter employed for the experiment, uses a finite impulse response (FIR) architecture and was implemented with FPGA (Xilinx XC6SLX9) with a discretized basis filter function (see Supplementary Materials for details). The same matched filter coefficient values were used for all results shown in Figs. 3-5. To adopt the matched filter for different sweep rates, its clock frequency was proportionally changed so that the basis function and the measured $V_{if}$ signals had the same scale.

**Acknowledgments**

This work was supported in part by the ERC Grant MAGICAL (N°669204), the U.S. National Science Foundation (Grants # EFMA-1641989 and # ECCS-1708982), by the Air Force Office of Scientific Research under the MURI grant # FA9550-19-1-0307, and by the Oakland University Foundation. V. I. acknowledges financial support from the EC grant GREAT (687973), and P.S. acknowledges financial support from the Enhanced Eurotalent programme of CEA as well as the French space agency CNES.

**Author contributions**

A.L. and U.E. conceived the experiments; A.L. designed the matched filter and the experiment; A.L., V.I. and P.S. performed the measurements and analyzed the data; S.L., J.L., V.T. and A.S. performed theoretical calculations; A.J. and R.F. prepared the samples; A.L., U.E. and A.S. wrote the manuscript with help from S.L.; A.S. and U.E. managed the project; all authors contributed to the manuscript, the discussion and analysis of the results.

**Competing interests**

The authors have no competing financial interests.

# Supplementary Material for the manuscript entitled

"Experimental demonstration of a rapid sweep-tuned spectrum analyzer with temporal resolution based on a spin-torque nano-oscillator" by

A. Litvinenko[1,*], V. Iurchuk[1], P. Sethi[1], S. Louis[2], V. Tiberkevich[2], J. Li[2], A. Jenkins[3], R. Ferreira[3], B. Dieny[1], A. N. Slavin[2], U. Ebels[1]

[1]SPINTEC – Univ. Grenoble Alpes, CEA, CNRS, IRIG – Grenoble – France

[2]Oakland University – Rochester – USA

[3]International Iberian Nanotechnology Laboratory (INL) – Braga – Portugal

* corresponding author: artem.litvinenko@cea.fr

## I. Characterization of the free running STNO properties

Representative results obtained on a vortex-state spin torque nano-oscillator (STNO) device for the free-running frequency, output power and FWHM linewidth ΔF vs. voltage are shown in Fig. S1, as measured using a spectrum analyzer. In this example, steady-state oscillations appear when the applied voltage is above $V = 0.2$V.

For the demonstration of the STNO performance as a spectrum analyzer (SA) the operational voltage was limited to a range where the STNO showed the best linewidth. For the device in Fig. S1 this is V ≈ 0.25 - 0.37V corresponding to an operational frequency range of $\Delta F_{sw} = 20$ MHz (between the frequencies $f$=290-310 MHz), with the FWHM generation linewidth around $2\Delta F$=250kHz and the output power of about $P_{out}$=0.25μW.

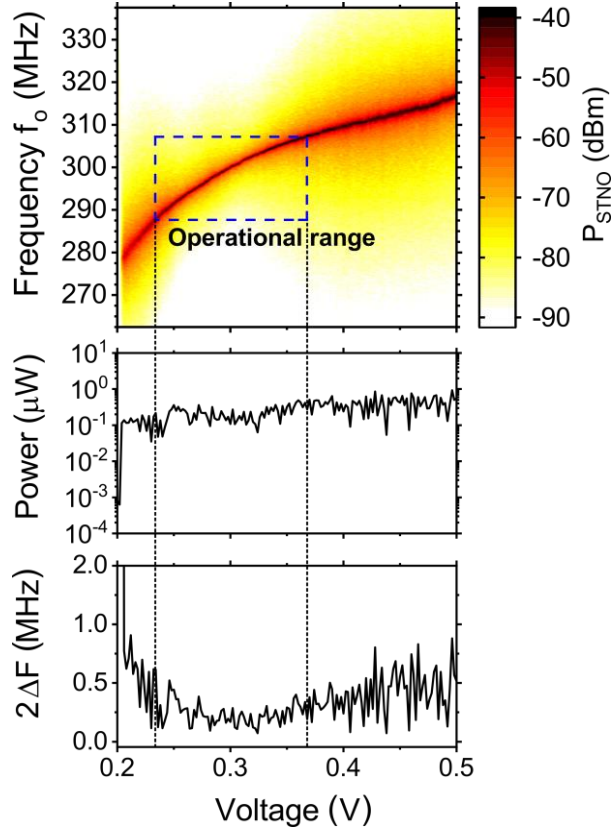

Figure S1: Free-running STNO properties vs. driving voltage: (a) frequency $f_o$; (b) output power; and (c) full width at half maximum linewidth $2\Delta F$.

## II. Compensation of the non-linear frequency-voltage dependence

As it can be seen from Fig. S1 and from Fig. 2 of the main text, the voltage-frequency dependence in the used STNO is strongly nonlinear. When a linear saw-tooth voltage signal $V_{sw}$, is applied to an STNO the response of the instantaneous frequency $f_o(t)$ generated by STNO also varies non-linearly in time as shown in Fig. S2(Ia).

The resulting non-linearity of the frequency-time dependences $V_{ref}$ and $V_{if}$ lead to the distortions in the shape and position of the resulting peak at the output of the matched filter, in particular at the points in time where the magnitude of the slope $df_o/dt$ differs substantially from the value at the center of the operation frequency range, for which the matched filter was set up.

As an example in Fig. S2I the voltage at the matched filter output is shown for four different input signal frequencies (panels I(b) – I(e)). Here only the signal in the panel I(c) is well resolved, while for all the other frequencies the output signal shows a broad peak or multiple peaks. Evidently these distortions spoil the frequency resolution $\Delta F_{RBW}$, and the device ability to accurately detect the frequency of the input signal.

Essentially this would mean that in such a nonlinear case a single matched filter would not be able to resolve and detect signals in the whole range of scanned frequencies. Thus, rather than using different matched filters with different time scales adapted to each instantaneous magnitude of the slope $df_o/dt$, it is preferable to linearize the dependence of the generated frequency on time.

To achieve a linear dependence $f_o(t)$ in our experiment, the natural non-linear (convex) dependence $f_o(V)$ was compensated by using a corresponding non-linear (concave) voltage sweep function $V_{sw}(t)$ as shown in Fig. S2, panel IIa.

Here the instantaneous STNO frequency $f_o(t)$ varies linearly with time, and the signals of all the four frequencies (panels II(b) – II(d)) are well resolved with the same frequency resolution $\Delta F_{RBW}$, and the same peak shape.

In order to adjust the shape of $V_{sw}$ resulting in the linear $f_o(t)$ dependance, a look-up table of the frequency-voltage-time dependence $\{f_o \mid V_{sw} \mid t\}$ was created using the data of Fig. S2 panel Ia. Then, the table is rearranged and interpolated to obtain a dependence $f_o(V_{sw}(t))$ with equidistant frequency intervals. This operation results in the generation of a non-linear time dependence of $V_{sw}(t)$, that is applied to the STNO. This method of compensation of the natural STNO nonlinearity leads to a linear correspondence between time and the STNO-generated frequency (see panel II(e) bottom scale).

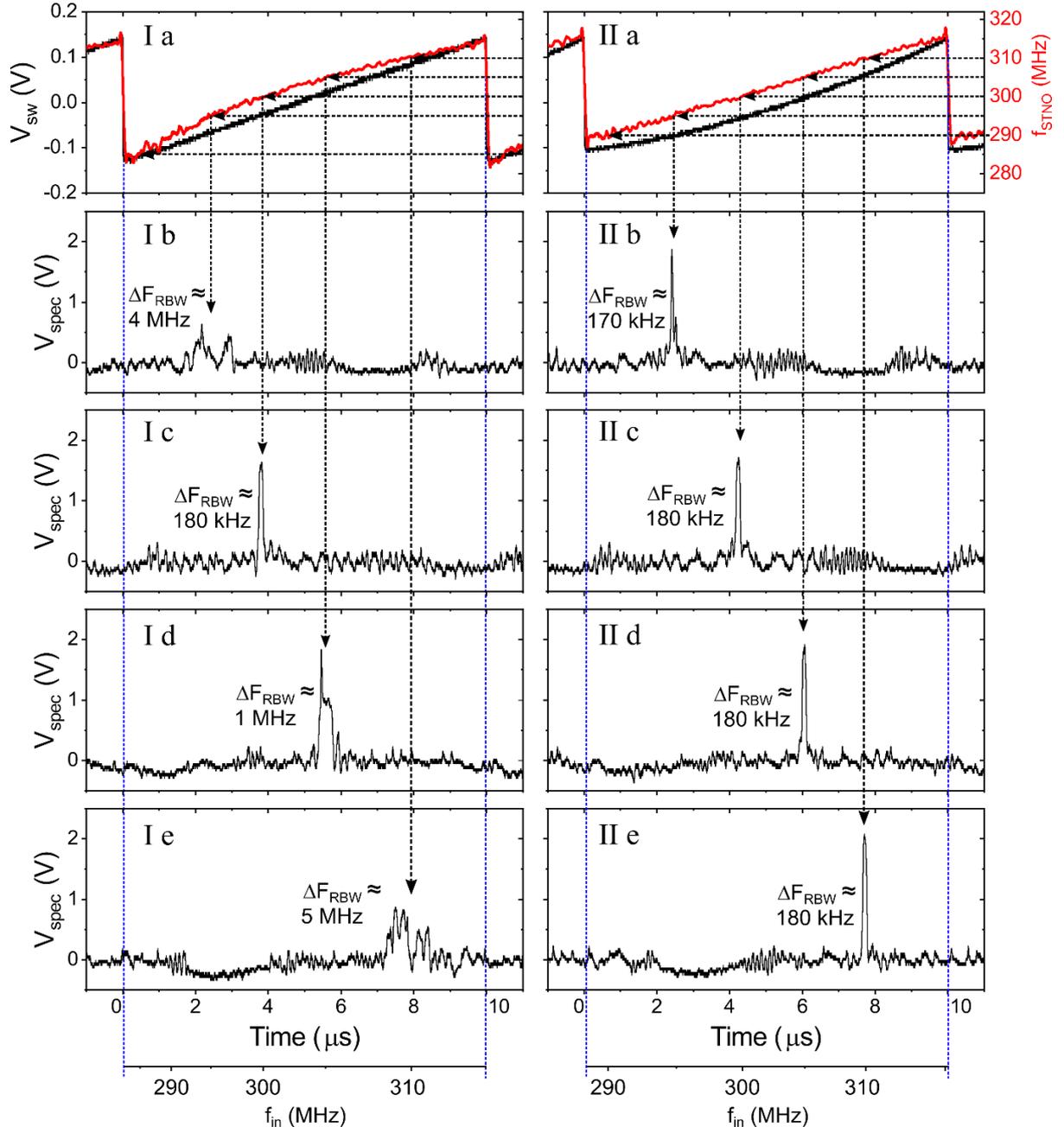

Figure S2. Compensation of non-linearity of the voltage-frequency dependence of the used vortex-state STNO: (Ia) A linear voltage sweep $V_{sw}$ generates a non-linear instantaneous frequency sweep $f_o(t)$ and (IIa) a non-linear voltage sweep $V_{sw}$ generates a linear instantaneous frequency sweep $f_o(t)$. In panels Ib-Ie and IIb-IIe the matched filter output $V_{spec}$ is shown for single tone signals of different frequencies (b) 295MHz, (c) 300MHz, (d) 305MHz and (e) 310MHz.

### III. Matched Filter

Matched filters are actively used in radar applications [S1], and can be implemented digitally using either Digital Signal Processors (DSP) or a Field Programmable Gate Array (FPGA). A matched filter performs a cross-correlation between a basis filter function and a noisy input signal. The resulting compressed peak has the form of a cross-correlation peak (see panels (v)-(vii) of Fig. 1(b) of the main text).

In our realization of a matched filter, a direct configuration with a finite impulse response (FIR) was chosen, see Fig. S3(a), since it is more hardware-efficient and preferred for DSP and FPGA implementation. The corresponding basis filter function of the matched filter is shown in Fig. S3(b) and was defined through digitized coefficients of the values $k_i$. These coefficient values $k_i$ are determined as follows: a continuous noise-free chirped signal $V_{if}(t)$ is generated using a behavioral model for a single-tone input signal $V_{in}$ whose frequency lies at the center of the sweeping range. The digitized data are transformed via FFT into the frequency domain, and then conjugated and transformed back into the time domain using the inverse fast Fourier transform (FFT):

$$k_i = IFFT\left[conj\left(FFT(V_{IF}(i))\right)\right]. \qquad (S1)$$

The values of the coefficients $k_i$ are, then, normalized, and are used as multiplication factors in the FIR architecture of the matched filter. In such a way the FIR filter with the coefficient values $k_i$ is matched to the digitized signal $V_{IF}(i)$.

In the STNO-SA experiment, the mixed signal $V_{if}(t)$ (see frame (v) of Fig. 1(b) in the main text) has the form of a symmetrical chirped wavelet when the input signal has a single frequency that lies at the center of the operational range of the STNO. The matched filter is optimized for this symmetrical chirp wavelet. However, the realized matched filter will also compress asymmetrical input chirp signals (i.e. signals whose frequency is not at the center of the operational range) for which there is only a partial correlation between the input signal and the basis function.

In order to get the highest possible SA resolution at lowest possible clock frequency of a FIR matched filter, the digitization of $V_{IF}(i)$ was done close to its Nyquist frequency.

Since for each sweeping frequency the time scale of the input chirp signal is different, the sampling frequency of the matched filter was adapted accordingly so that the input digitized signal has the same shape for which the matched filter was optimized.

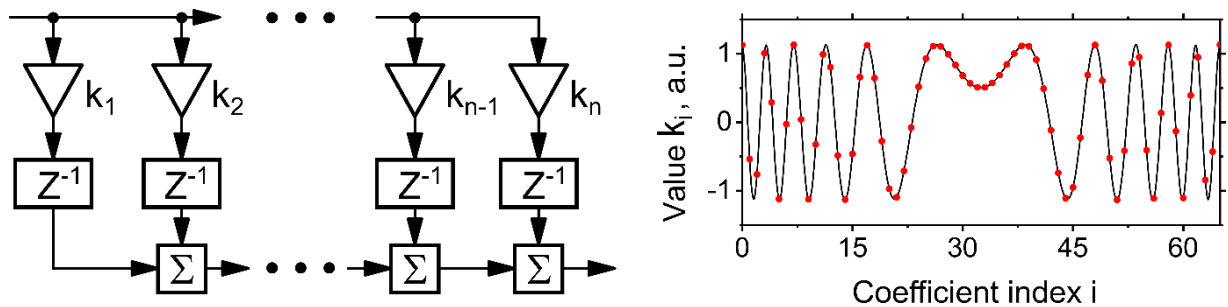

Fig. S3: (a) Direct FIR architecture of a matched filter. (b) The coefficient values $k_i$ of the matching basis filter function

**Reference**

[S1] Turin, G. L. "An introduction to matched filters". *IRE Transactions on Information Theory*. **6** (3): 311–329. (1960).